\documentclass{emulateapj}
\usepackage{bm}

\newcommand{\EQ}{\begin{equation}}
\newcommand{\EN}{\end{equation}}
\newcommand{\EQA}{\begin{eqnarray}}
\newcommand{\ENA}{\end{eqnarray}}

\newcommand{\Sec}[1]{\S\ref{#1}}

\newcommand{\Fig}[1]{Fig.~\ref{#1}}

\newcommand{\Tab}[1]{Table~\ref{#1}}

\newcommand{\bra}[1]{\langle #1\rangle}

{}

\newcommand{\meanBB}{\overline{\bm{B}}}
\newcommand{\meanJJ}{\overline{\bm{J}}}
\newcommand{\meanUU}{\overline{\bm{U}}}
\newcommand{\meanWW}{\overline{\bm{W}}}

{}
{}
{}
{}
{}
{}
{}

\newcommand{\meanB}{\overline{B}}

\newcommand{\nnn}{\hat{\mbox{\boldmath $n$}} {}}

\newcommand{\UU}{{\bm{U}}}

\newcommand{\uu}{{\bm{u}}}
\newcommand{\BB}{{\bm{B}}}

\newcommand{\JJ}{{\bm{J}}}

\newcommand{\AAA}{{\bm{A}}}

\newcommand{\ff}{\mbox{\boldmath $f$} {}}

\newcommand{\FF}{{\bm{F}}}

\newcommand{\nab}{\mbox{\boldmath $\nabla$} {}}
\newcommand{\OO}{\mbox{\boldmath $\Omega$} {}}
\newcommand{\oo}{\mbox{\boldmath $\omega$} {}}

\newcommand{\SSSS}{\mbox{\boldmath $\sf S$} {}}

\newcommand{\DD}{{\rm D} {}}

\newcommand{\dd}{{\rm d} {}}
\newcommand{\const}{{\rm const}  {}}

\def\half{{\textstyle{1\over2}}}

\def\onethird{{\textstyle{1\over3}}}

\newcommand{\G}{\,{\rm G}}

\newcommand{\nHz}{\,{\rm nHz}}

\newcommand{\kG}{\,{\rm kG}}

\newcommand{\mpers}{\,{\rm m/s}}

\newcommand{\Mm}{\,{\rm Mm}}
\newcommand{\Mx}{\,{\rm Mx}}

\newcommand{\yr}{\,{\rm yr}}

\newcommand{\yapj}[3]{ #1, {ApJ,} {#2}, #3}

\newcommand{\yapjl}[3]{ #1, {ApJ,} {#2}, #3}

\newcommand{\yan}[3]{ #1, {AN,} {#2}, #3}
\newcommand{\yzfa}[3]{ #1, {Z.\ f.\ Ap.,} {#2}, #3}

\newcommand{\yana}[3]{ #1, {A\&A,} {#2}, #3}

\newcommand{\yanar}[3]{ #1, {A\&AR,} {#2}, #3}

\newcommand{\ygafd}[3]{ #1, {Geophys. Astrophys. Fluid Dyn.,} {#2}, #3}

\newcommand{\yjfm}[3]{ #1, {JFM,} {#2}, #3}

\newcommand{\ypp}[3]{ #1, {Phys. Plasmas,} {#2}, #3}
\newcommand{\ysov}[3]{ #1, {Sov. Astron.,} {#2}, #3}

\newcommand{\yjetp}[3]{ #1, {Sov. Phys. JETP,} {#2}, #3}

\newcommand{\yaraa}[3]{ #1, {ARA\&A,} {#2}, #3}

\newcommand{\yprl}[3]{ #1, {PRL,} {#2}, #3}

\newcommand{\yptrs}[3]{ #1, {Phil. Trans. Roy. Soc.,} {#2}, #3}
\newcommand{\ymn}[3]{ #1, {MNRAS,} {#2}, #3}
\newcommand{\ynat}[3]{ #1, {Nat,} {#2}, #3}
\newcommand{\ysci}[3]{ #1, {Sci,} {#2}, #3}
\newcommand{\ysph}[3]{ #1, {Solar Phys.,} {#2}, #3}

\newcommand{\ypre}[3]{ #1, {PRE,} {#2}, #3}

\newcommand{\yjour}[4]{ #1, {#2}, {#3}, #4}

\newcommand{\ybook}[3]{ #1, {#2} (#3)}
\newcommand{\yproc}[5]{ #1, in {#3}, ed. #4 (#5), #2}

\newcommand{\pan}[1]{ #1, {AN,} (in press)}

\submitted{}
\graphicspath{{./fig/}{./png/}}
\journalinfo{The Astrophysical Journal, 625, 2005 June 1}

\received{2004 December 15}
\accepted{2005 February 11}
\begin{document}
\title{The case for a distributed solar dynamo shaped by near-surface shear}
\author{Axel Brandenburg}
\affil{Isaac Newton Institute for Mathematical Sciences,
20 Clarkson Road, Cambridge CB3 0EH, UK; and\\
Nordita, Blegdamsvej 17, DK-2100 Copenhagen \O, Denmark;
\url{brandenb@nordita.dk}\\
}

\begin{abstract}
Arguments for and against the widely accepted picture of a solar dynamo being
seated in the tachocline are reviewed and alternative ideas concerning dynamos
operating in the bulk of the convection zone, or perhaps even
in the near-surface shear layer, are discussed.
Based on the angular velocities of magnetic tracers
it is argued that the observations are compatible with
a distributed dynamo that may be strongly shaped by the
near-surface shear layer.
Direct simulations of dynamo action in a slab with turbulence and
shear are presented to discuss filling factor and tilt angles of
bipolar regions in such a model.
\end{abstract}


\section{Introduction}

There appears to be general consensus that the solar magnetic field is generated
and stored in the overshoot layer near the bottom of the convection zone
(Spiegel \& Weiss 1980, Golub et al.\ 1981, Galloway \& Weiss 1981,
van Ballegooijen 1982, Choudhuri 1990).
This layer is now believed to coincide with the tachocline,
i.e.\ the radial shear layer at the bottom of the convection zone where the
latitudinal differential rotation changes into rigid rotation in the
radiative zone (Spiegel \& Zahn 1992).
The main arguments in favor of this proposal are connected with flux
storage over times long enough for the shear to amplify the toroidal field
(Moreno-Insertis, Sch\"ussler, \& Ferriz-Mas 1992),
and with the observed size of active regions ($\sim100\Mm$) being comparable
with the typical eddy scale at the bottom of the convection zone
(Galloway \& Weiss 1981), as well
as the observed fidelity of Hale's polarity law;
see Fisher et al.\ (2000), Tobias (2002), Sch\"ussler (2002),
Ossendrijver (2003), Fan (2004), and Weiss (2005) for recent reviews.
All these aspects are intimately related to the thin flux tube picture.
Indeed, one of the big successes of the thin flux tube approximation 
(Spruit 1981, Moreno-Insertis 1986, Chou \& Fisher 1989) is the
quantitative prediction of Joy's law describing the latitudinal dependence
of the observed tilt angles of bipolar regions.
It is found that Joy's law is obeyed only for
flux tubes with magnetic fields that are of the order of $10^5\G$
(D'Silva \& Choudhuri 1993, Sch\"ussler et al.\ 1994,
Caligari, Moreno-Insertis, \& Sch\"ussler 1995).
This result poses rather stringent demands on dynamo theory that
are hard to meet.
Although it may already be hard for the differential rotation in the tachocline
to amplify a poloidal field to a strength of $\sim10^5\G$, which may require
flux intensification by exploding flux tubes (Rempel \& Sch\"ussler 2001),
it is not obvious how to explain the production of strong and
sufficiently coherent {\it poloidal} field that is necessary to
produce the toroidal field.

The purpose of the present paper is to discuss the difficulties for
dynamo theory in meeting these demands and to reconsider the alternative
scenario that the solar dynamo may operate in the bulk of the convection zone,
or perhaps even in the near-surface shear layer in the upper $35\Mm$ of the sun.
This is the layer where recent helioseismological inversions have shown
marked negative radial shear (Howe et al.\ 2000a,
Corbard \& Thompson 2002, Thompson et al.\ 2003).
The presence of a deeper layer that spins about 5\% faster than the
photosphere has always been anticipated based on the higher rotation
rate of magnetic tracers (Gilman \& Foukal 1979, Golub et al.\ 1981).
It remained unclear, however, just how deep or shallow this layer really is.
The most natural assumption {\it at the time} was to place this layer near the
bottom of the convection zone where magnetic buoyancy is weak and shear
could work on the field unimpeded by the turbulence.
In the early days of mean field dynamo theory a negative radial $\Omega$
gradient was already anticipated (Stix 1976, Yoshimura 1976) because it
would explain the observed anticorrelation of the signs of mean azimuthal
field (inferred from the orientation of bipolar regions) and the mean
radial field (measured by magnetograms).
This is because negative radial shear turns a positive radial field into a
negative azimuthal field, producing the observed anticorrelation.
A negative radial gradient of angular velocity seemed confirmed by
observations of very young sunspots that rotate faster than older ones
(Tuominen 1962), suggesting that they may be anchored in the layer where
the angular velocity is maximum (Tuominen \& Virtanen 1988,
Balthasar, Sch\"ussler, \& W\"ohl 1982, Nesme-Ribes, Ferreira, \& Mein 1993,
Pulkkinen \& Tuominen 1998).

The sunspot observations are not easily explained by interface dynamos,
unless one is able to show that the angular velocity of magnetic
tracers is just the pattern speed of a traveling wave phenomenon.
A similar proposal has been made to explain
the observed pattern speed of the supergranulation
(Gizon, Duvall, \& Schou 2003, Schou 2003, Busse 2004).

The importance of the near-surface shear layer 
has already been investigated by Dikpati et al.\ (2002)
who studied the effects of near-surface radial shear
on a flux transport dynamo.
They came to the conclusion that the effect of the near-surface shear
layer is subdominant in the context of the flux transport model studied
earlier by Dikpati \& Charbonneau (1999).
Mason, Hughes, \& Tobias (2004) also considered the issue of near-surface
dynamo action in the context of a two-layer dynamo (one at the top and
one at the bottom of the convection zone).
They allowed for an additional $\alpha$ effect in the upper
layers, retaining only the radial shear in the tachocline.
They came to the conclusion that the near-surface dynamo was harder to
excite due to the larger distance to the tachocline.

\begin{figure}[t!]\begin{center}
\plotone{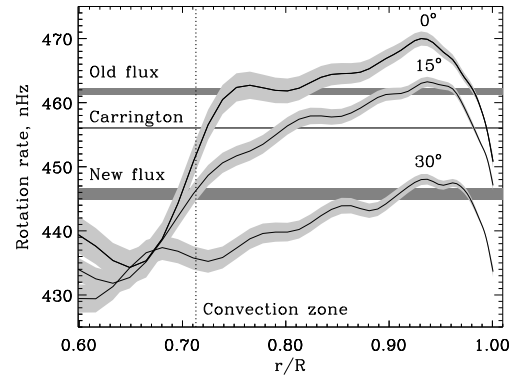}
\end{center}\caption[]{
Radial profiles of the internal solar rotation rate, as inferred from
helioseismology.
The angular velocities of active zones at the beginning of the cycle
(at $\approx30^\circ$ latitude) and near the end (at $\approx4^\circ$)
is indicated by horizontal bars, which intersect the profiles of angular
velocity at $r/R\approx0.97$.
[Adapted from Benevolenskaya et al.\ (1999).]
}\label{howe00}\end{figure}

There are several arguments in favor of a dynamo operating in or being
strongly controlled by the near-surface layer of the sun.
First, in the outer $35\Mm$ (corresponding to fractional radii
$r/R\ge0.95$) the negative radial shear, together
with an $\alpha$ effect of the usual sign, would easily
explain the observed equatorward migration of sunspot activity
(e.g.\ Parker 1979).
Second, the negative phase relation between radial and azimuthal mean fields,
$\meanB_r$ and $\meanB_\phi$, respectively,
would be automatically satisfied (Stix 1976, Yoshimura 1976).
Third, the radial near-surface shear is particularly strong between the equator
and $30^\circ$ latitude, and weak near the poles; see \Fig{howe00}.
In the tachocline, by contrast, there is hardly any radial shear 
at $30^\circ$ latitude and maximum shear near the poles.
Invoking near-surface shear for producing sunspot activity
would naturally explain the strongest production of sunspot
activity at low latitudes and the much weaker magnetic activity at
high latitudes with a possible poleward migration (Stix 1974),
provided the near surface shear changes sign at high latitudes,
as is perhaps indicated by helioseismology (Thompson et al.\ 2003).
[The poleward branch could also be due poleward flows -- as demonstrated
convincingly by flux transport models; see Baumann et al.\ (2004)
and Sch\"ussler (2005).]
Fourth, the rotational velocity of very young sunspots (age less than
1.5 days) is $14.7^\circ/{\rm day}$ at low latitudes
(Pulkkinen \& Tuominen 1998), corresponding to $473\nHz$, which is about
the largest angular velocity measured with helioseismology anywhere in
the sun; cf.\ \Fig{howe00}.
This corresponds to the angular velocity at a radius $r/R=0.95$,
which is $35\Mm$ below the surface.
Similar conclusions can be drawn from the apparent angular velocity
of old and new magnetic flux at different latitudes
(Benevolenskaya et al.\ 1999).
We will return to these observations in \Sec{NearSurface}.

The main argument against distributed and near-surface shear layer
dynamos is that magnetic
flux tubes are highly buoyant in the convection zone proper (Parker 1975).
Thus, too much magnetic flux may be lost, on time scales so short that the
shear cannot amplify the poloidal field to significant field strengths.
However, over the past 15 years simulations have shown that magnetic
buoyancy is strongly offset by the action of turbulent pumping,
which leads to a net accumulation of
magnetic energy at the bottom of the convection zone (Nordlund et al.\
1992, Brandenburg et al.\ 1996, Tobias et al.\ 1998, 2001,
Dorch \& Nordlund 2001, Ossendrijver et al.\ 2002, Ziegler \& R\"udiger 2003).
Indeed, magnetic pumping is now also being invoked to keep down
the horizontal magnetic fields just outside the penumbra of sunspots
(Thomas et al.\ 2002, Weiss et al.\ 2004).
Thus, we can envisage pumping as a mechanism that tries to keep
the surface clean of magnetic fields, but it can do so only
approximately and only if the field is not already too strong.

Another argument against near-surface dynamos is the high degree
of turbulence in the upper layers, which could lead to
strong random distortions of the orientation of flux tubes.
This leads to the notion of rising flux tubes being
`brain washed' during their ascent (Sch\"ussler 1983, 1984),
i.e.\ they lose their original east-west orientation and
would not obey Hale's polarity law.
However, this picture derives originally from the idea that flux tubes
are produced in deep layers at or below the overshoot layer and are then
subjected to a more passive buoyant rise through the convection zone.
Here, however, we are envisaging the production of sunspots much closer to the
surface, so the notion of flux tubes rising through a major portion of
the convection zone is not invoked.
Indeed, local helioseismology suggests a picture quite compatible
with sunspots being a shallow surface phenomenon
(Kosovichev, Duvall, \& Scherrer 2000, Kosovichev 2002).
The actual sunspot formation might then be the result of convective
collapse of magnetic fibrils (Zwaan 1978, Spruit \& Zweibel 1979),
possibly facilitated by negative turbulent magnetic pressure effects
(Kleeorin, Mond, \& Rogachevskii 1996) or by
an instability (Kitchatinov \& Mazur 2000)
causing the vertical flux to concentrate into a tube.

It should be noted that the picture of shallow sunspots does not
necessarily contradict the idea of strong flux tubes rising to the
surface.
In fact, as the tube rises to the surface, it must eventually undergo
catastrophic expansion (Moreno-Insertis, Caligari, \& Sch\"ussler 1995).
This would detach the forming active region and its sunspots from its
roots (Schrijver \& Title 1999, Sch\"ussler 2005), which might then
be compatible with the shallow sunspot picture from local sunspot
helioseismology.

Yet another potential problem with near-surface shear layer dynamos are
the relatively short turbulent time scales.
However, in 35 Mm depth the typical turnover time is, according to
mixing length theory (Spruit 1974), already 1--3 days.
Therefore the inverse Rossby number, $2\Omega\tau$, is of the order
of unity, so the turbulence is certainly beginning to be affected by rotation.
As is familiar from mean field dynamo theory, the combination of poloidal
and toroidal fields really corresponds to a right-handed spiral in the
northern hemisphere.
Thus, whenever parts of this spiral touch the surface
they produce a bipolar active region with the observed tilt angle.
This will be discussed further in \Sec{BipolarRegions}.

We now turn to the question of small scale magnetic fields.
For magnetic Prandtl numbers as small as those in
the sun, the magnitude of the turbulent magnetic fields
from local small scale dynamo action at the top of the convection zone
(Cattaneo 1999) is possibly not much stronger than the magnitude of
fields from the large scale dynamo.
This suggestion is motivated by the recent realization
that small scale dynamo action
(as originally explored by Kazantsev 1968) becomes
either completely impossible or at least
much harder to excite when the magnetic
Prandtl number becomes small (Schekochihin et al.\ 2004a,
Boldyrev \& Cattaneo 2004, Haugen, Brandenburg, \& Dobler 2004).
The dominance of small scale over large scale dynamo activity
in global simulations of solar-like convection
(Brun, Miesch, \& Toomre 2004) might therefore also be
related to the fact that the magnetic Prandtl number
is not small in those simulations.

Another important aspect is the fact that in the presence
of shear, turbulent dynamos can produce and maintain fields of
equipartition strength (Brandenburg et al.\ 2005).
We will discuss some of those models also in \Sec{Simulations}.
Although such models still lack important aspects of solar dynamos
(convection, stratification, and rotation), they are quite suitable
for testing new effects in mean field theory,
for example the shear--current effect
(Rogachevskii \& Kleeorin 2003, 2004) and current helicity fluxes
(Brandenburg \& Sandin 2004, Subramanian \& Brandenburg 2004).
We begin by discussing first the problems associated with the current
picture of dynamos operating in the tachocline.
A summary of arguments discussed in the text is given in
\Tab{ArgumentsSummary}.

\begin{deluxetable*}{lll}
\tabletypesize{\scriptsize}
\tablecaption{Summary of arguments for and against tachocline and
distributed dynamos discussed in the text}
\tablehead{ arguments & tachocline dynamos & distributed/near-surface dynamos}
\startdata
\label{ArgumentsSummary}
in favor
& flux storage
   & negative surface shear yields equatorward migration\\
& turbulent distortions weak
   & correct phase relation \\
& correct butterfly diagram with meridional circulation
   & strong surface shear at latitudes where the spots are \\
& size of active regions ($\sim100\Mm$) naturally explained
   & $\max(\Omega)/2\pi=473\nHz$ agrees with $\Omega(\mbox{youngest spots})$ \\
& 
   & active zones move with $\Omega(0.95)$ \\
& 
   & $11\yr$ variation of $\Omega$ seen in the outer $70\Mm$ \\
& 
   & even fully convective stars have dynamos \\
\\
against
& $100\kG$ field hard to explain
   & strong turbulent distortions
\\
& flux tube integrity during ascent
   & rapid buoyant losses \\
& too many flux belts in latitude
   & too many flux belts if dynamo only in shear layer\\
& maximum radial shear at the poles
   & not enough time for shear to act \\
& no radial shear where sunspots emerge
   & long term stability of active regions\\
& quadrupolar parity preferred
   & profile of $\Omega(\mbox{youngest})$ by $4\nHz$ above $\Omega(0.95)$\\
& wrong phase relation
   & possible anisotropies in supergranulation \\
& $1.3\yr$ instead of $11\yr$ variation of $\Omega$ at base of CZ
   & \\
& coherent meridional circulation pattern required
   & \\
\enddata
\end{deluxetable*}

\section{Problems with tachocline dynamos}

By tachocline dynamos we mean dynamos where the main shear that is
responsible for the cyclic toroidal fields originates from the tachocline.
These dynamos take into account the measured differential rotation
profile, although sometimes the latitudinal shear is neglected
(e.g.\ Parker 1993, Choudhuri, Sch\"ussler, \& Dikpati 1995).
Tachocline dynamos can be divided into three main subclasses:
(i) overshoot dynamos where there is only a negative $\alpha$ effect
in the overshoot layer, (ii) interface dynamos where a negative $\alpha$ effect
is assumed in the upper parts of the convection zone, and
(iii) Babcock-Leighton type flux transport dynamos where the
$\alpha$ effect is also located near the surface, but it is now
positive and there is meridional circulation
transporting flux in the overshoot layer from high latitudes
toward the equator.

One of the longstanding problems with dynamos operating in a thin layer
at the bottom of the convection zone is the large number of oppositely
oriented toroidal flux belts in each hemisphere
(Moss, Tuominen, \& Brandenburg 1990).
This tends to produce a rather unrealistic butterfly diagram.
This problem can partly be alleviated by increasing the thickness
of the overshoot layer (R\"udiger \& Brandenburg 1995) to about
$50\Mm$, which is beyond the currently accepted thickness of
the overshoot layer of about $7\Mm$ or less (Basu 1997).

Another problem is the strong radial shear at polar latitudes
in the tachocline.
This leads to a dominance of magnetic activity in polar regions.
It is therefore customary to postulate an artificially modified latitudinal
dependence of the $\alpha$ effect.
R\"udiger \& Brandenburg (1995) assumed that $\alpha$ was
proportional to $\cos^2\!\theta$, where $\theta$ is colatitude,
and Markiel \& Thomas (1999) assumed that $\alpha$ was proportional to
a gaussian concentrated around the equator.
This manipulation was originally motivated by the possible presence
of higher order terms quantifying the combined influence of stratification
and rotation.
Simulations of Ossendrijver et al.\ (2002) have indeed confirmed a
suppression of $\alpha$ near the poles.
Nevertheless, it remains
puzzling that at $30^\circ$ latitude, where sunspots first emerge, the
radial shear in the tachocline basically vanishes.
So there should not be any local toroidal field generation.
This problem may however be alleviated in the context of flux transport
dynamos, as will be discussed at the end of this section.

We recall that the {\it positive} radial angular velocity gradient
in the tachocline stretches a positive $\meanB_r$ into a positive
$\meanB_\phi$, so that $\meanB_r\meanB_\phi$ is also {\it positive}.
As discussed in the introduction, this is in conflict with observations
(Yoshimura 1976, Stix 1976).
Although in some models $\meanB_r\meanB_\phi$ can still be negative during
certain intervals and in certain latitudes, this cannot be regarded as a
robust or well understood feature.
Also, the occasional intervals of positive $\meanB_r\meanB_\phi$
seen in some models (K\"uker, R\"udiger, \& Schultz  2001) depend on
assumptions about the depth were the toroidal field is evaluated.
Furthermore, these models rely on the negative $\alpha$ that is expected
at the bottom of the convection zone (Yoshimura 1972, Krivodubskii 1984,
R\"udiger \& Kitchatinov 1993).

If the cyclic field in the sun really originates from the tachocline,
one would expect to see cyclic modulations of the local angular velocity,
similar to those seen at the surface (Howard \& LaBonte 1980).
Instead, there is possible evidence for a shorter $1.3\yr$ period at
the base of the convection zone (Howe et al.\ 2000a).
This would suggests that the field responsible for the $22\yr$ cycle cannot
come from the tachocline, but rather from the outer $70\Mm$ of the sun
where $11\yr$ variations have indeed been seen (Howe et al.\ 2000b,
Vorontsov et al.\ 2002).
Indeed, a recent model by Covas et al.\ (2001) explains the $1.3\yr$
period in terms of spatio-temporal fragmentation, where the dynamo has a
shorter period at the bottom of the convection zone and a longer period
in the upper parts.
In this model the field responsible for the $22\yr$ cycle would originate
from the upper parts of the convection zone (CZ).

We note in passing that, if a tachocline was really crucial for a dynamo
to work, one might expect a break in the magnetic activity toward late
M dwarfs that become fully convective and therefore lack a tachocline.
This is not observed (Vilhu 1984).
However, this argument is not really compelling because
it can be argued that the fields of fully convective stars are
only of small scale (Durney, De Young, \& Roxburgh 1993).

Finally, a problem with any model drawing its field from deep underneath
is that it is not easy to imagine that a flux tube can maintain its
integrity while rising over 20 pressure scale heights from
the bottom of the convection zone to the top.
Indeed, direct simulations show that a large amount of twist is needed
to keep the tubes intact over at least a few pressure scale heights
(Moreno-Insertis \& Emonet 1996).
On the other hand, although a modest amount of twist can be useful
for explaining the so-called $\delta$ spots (Fan et al.\ 1999),
too much twist can make the tubes kink-unstable
(e.g., Linton, Longcope, \& Fisher 1996).
It is also not clear how the faster sunspot proper motion of very young
sunspots (Pulkkinen \& Tuominen 1998) can be explained if the spots were
rooted in the tachocline.
It would be much more straightforward if they were rooted near
the maximum of $\Omega$ about $35\Mm$ below the surface.

A different class of models are the Babcock-Leighton type flux transport
dynamos, as recently studied by Dikpati \& Charbonneau (1999),
Nandy \& Choudhuri (2002), and Chatterjee et al.\ (2004).
Here the $\alpha$ effect is assumed to come from the surface layers,
so $\alpha$ would be positive.
These models deal with some of the aforementioned problems by
invoking a grand meridional circulation pattern playing the role of a
conveyor belt that transports flux through the tachocline from high
latitudes to the equator.
This circulation is responsible for driving the dynamo waves equatorward
and are also determining the cycle period (Durney 1995,
Choudhuri et al.\ 1995, K\"uker et al.\ 2001).
A polar branch, on the other hand, can be explained by postulating
a two-cell circulation pattern.
It is this type of model for which the effect of the near-surface shear
layer has recently been investigated by Dikpati et al.\ (2002).
However, form and magnitude of the meridional circulation in the
sun are quite uncertain.
Direct simulations by Miesch et al.\ (2000) suggest a rather more
irregular pattern of many cells changing with time.
If this result continues to persist also in more realistic simulations,
it would render the flux transport picture rather fragile.

\section{Problems with distributed dynamos}
\label{NearSurface}

We begin by discussing first the evidence
from magnetic tracers in favor of their near-surface anchoring
and turn then to the more theoretical arguments supporting
the notion of distributed dynamos that are being strongly
affected by the near-surface shear layer.

\begin{figure}[t!]\begin{center}
\plotone{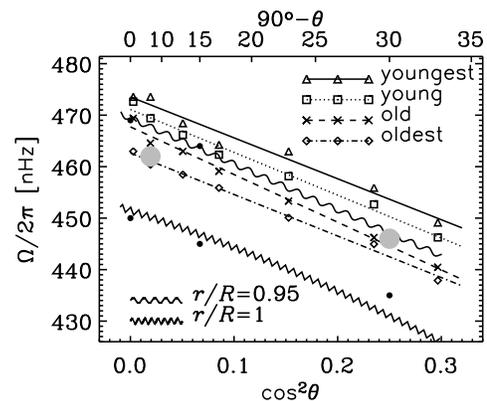}
\end{center}\caption[]{
Angular velocity of sunspots of different age as a function of colatitude
$\theta$ compared with the helioseismological internal angular velocity at
$r=0.95R$ and with the Doppler velocity at the surface.
The large gray dots at about $30^\circ$ and $4^\circ$ denote the angular
velocities of active zones respectively near the beginning and near
the end of the cycle.
The three small dots near the zig-zag line
are the helioseismological angular velocities at $r=0.7R$.
}\label{pdiffrot}\end{figure}

Magnetic tracers have long been known to rotate faster than
the photospheric plasma (e.g.\ Gilman \& Foukal 1979).
One possibility is that magnetic flux tubes possess an angular
velocity that is in excess of that of the surrounding plasma
(Wilson 1987).
Using the data of Pulkkinen \& Tuominen (1998) we plot
in \Fig{pdiffrot} the angular velocity of sunspots of different
age (from 1.5 days to several months) versus colatitude, where we have
fitted their angular velocity to the common representation
$\Omega=a+b\cos^2\theta$, which leads to a linear graph when plotted
versus $\cos^2\theta$.
These results are compared with the helioseismologically determined
angular velocity at $r=0.95R$ and $0.7R$, as well as with the
Doppler velocity at the surface, using a fit up to $\cos^4\theta$,
as quoted by Thompson et al.\ (2003).


Let us now discuss the properties of a dynamo operating in the bulk or
in the upper layers of the sun.
We envisage a dynamo that operates very much like a classical
$\alpha\Omega$ dynamo as it was proposed in the early days
of mean field dynamo theory.
In particular, one may anticipate a field that is not strongly fibril,
as is indeed confirmed by simulations of turbulent dynamos with shear
(Brandenburg, Bigazzi, \& Subramanian 2001).
Furthermore, in the uppermost $3\Mm$ of the sun turbulent downward
transport is far too strong to let any significant field appear
at the surface, except in active regions which emerge as the result
of strong flux segregation into strongly and weakly magnetized regions,
as demonstrated by large aspect ratio magneto-convection simulation
(Tao et al.\ 1998).

There are however several new problems.
Most important is the fact that the near surface shear layer is rather
thin, so we may have a problem of too many toroidal flux belts
if the dynamo was solely confined to the surface layer.
There are other possible problems where only a preliminary
discussion can be offered.
This includes potentially observable effects of near surface magnetic
activity on the supergranulation.
It is conceivable that the supergranulation may show significant
alignment with the mean field.
So far, no such anisotropy has been reported, although we do know that the
cell size of the normal surface granulation does change with the cycle
(Houdek et al.\ 2001).

The other aspect concerns Joy's law which has successfully been reproduced
within the framework of the thin flux tube approximation.
In the context of a distributed dynamo the inclination of bipolar regions
is primarily controlled by the sense of the latitudinal shear
(patches at higher latitudes lack behind those at lower latitudes).
There is also a contribution of $\alpha$ to the tilt (positive $\alpha$
produces positively helical fields whose interceptions with the surface
yield a solar tilt).
The latter effect is however subdominant, as is seen 
from a turbulence simulation (see the next section).

We should mention the phenomenon of active zones.
They constitute patches of recurrent magnetic activity over months
and sometimes years.
Recent investigations by Benevolenskaya et al.\ (1999) showed that
these patches have different angular velocity at different depths,
corresponding to the local angular velocity at radii between
$0.95\,R$ and $0.98\,R$, suggesting again that these magnetic
activity complexes are anchored within the near-surface shear layer;
see \Fig{howe00} and the gray dots in \Fig{pdiffrot}.
One may picture these activity complexes as more strongly magnetized
regions which can only decay slowly, possibly because of magnetic
helicity conservation of perhaps because magnetic fields can have
a tendency to segregate into strongly and weakly magnetized regions,
as is found in large aspect ratio magnetoconvection with imposed
field (Tao et al.\ 1998).

An important consideration for tachocline dynamos is whether
the observed emergent flux of $10^{24}\Mx$ over the full solar
cycle can be produced (Galloway \& Weiss 1981).
In the present context we are thinking of mean toroidal fields of the
order of $300\G$, which is about one tenth of the strength of the
mean field usually envisaged for the overshoot layer.
However, because of the larger cross-sectional surface of, say,
$(200\Mm)^2$, this will still produce the required $10^{24}\Mx$.

Global simulations of the solar dynamo are becoming more advanced.
The models of Brun et al.\ (2004) show distributed dynamo
action of the type envisaged in this paper.
However, these models have some characteristic properties that are
different from the observed solar field.
Most important is perhaps the relatively large ratio of poloidal to
toroidal field, which suggests that the effect of the differential
rotation is not sufficiently prominent or, conversely, the effect of
the small scale turbulent field is too prominent.
In the following we discuss a possible cause of this and address the
question how this may change with increasing resolution and larger fluid
and magnetic Reynolds numbers.

There are two distinct properties of turbulent dynamo action that
seem to depend on microscopic viscosity and diffusivity.
First, the magnetic field in the simulations may be dominated by small
scale dynamo action (where helicity and shear are unimportant).
At small magnetic Prandtl number the small scale dynamo is much harder
to excite (Schekochihin et al.\ 2004a, Boldyrev \& Cattaneo 2004,
Haugen et al.\ 2004) and may become subdominant,
allowing the large scale dynamo effect to become more prominent.
Second, at high magnetic Reynolds numbers the large scale dynamo
time scale tends to be constrained by magnetic helicity conservation
(Brandenburg 2001).
Small scale magnetic helicity fluxes throughout the domain become
important to allow -- and even facilitate -- large scale dynamo action
(Blackman \& Field 2000, Kleeorin et al.\ 2000, 2003, Vishniac \& Cho 2001,
Subramanian \& Brandenburg 2004, Brandenburg \& Sandin 2004).
The convective dynamos in simulations of Brun et al.\ (2004),
for example, generate only a weak mean field.
This raises the possibility that such dynamos are of a different type
than the large scale dynamo that is believed to operate in the sun.
For this reason we now inspect a somewhat different simulation that
lacks convection and stratification, but where there is sufficient
shear to generate a prominent large scale field.
We use this simulation to find some guidance regarding the question
of how fibril the field of the sun is and whether it might provide
the right conditions for bipolar regions to form.

\section{Guidance from simulations}
\label{Simulations}

The existence of fibril fields (Parker 1982) is crucial in a scenario
where strong flux tubes rise through the convection zone to form sunspot
pairs at the surface.
The fibril nature of the field is also the main reason why
magnetic buoyancy may be so important.
Fibril fields have indeed been seen in simulations of
forced and convective turbulence
where mostly a small scale dynamo is in operation
(Nordlund et al.\ 1992, Politano, Pouquet, \& Sulem 1995,
Brandenburg et al.\ 1996).
However, this picture changes in the presence of strong shear, for example
in the case of accretion disc turbulence (Brandenburg et al.\ 1995)
or in the case of imposed shear (Brandenburg et al.\ 2001).
We begin with a brief description of the model and then discuss whether
the field is fibril and whether it is able to form bipolar regions.

\subsection{Description of the model}

The model discussed in this paper is basically equivalent to the model
studied recently by Brandenburg \& Sandin (2004), except that no external
field is imposed.
In this model the turbulence is driven by a forcing function $\ff$
that consists of eigenfunctions of the curl operator (with wavenumbers
$4.5\leq k_{\rm f}\leq5.5$) and of a large scale component with wavenumber
$k_1=1$.
The domain is of size $\half\pi\times2\pi\times\half\pi$, representing a
cartesian approximation $(x,y,z)$ to a sector in the sun between $0^\circ$
and $30^\circ$ latitude.
Thus, $(x,y,z)$ corresponds to $(r,\phi,-\theta)$, where $\phi$ is
longitude and $\theta$ colatitude.
The forcing function is arranged such that a mean flow of the form
\begin{equation}
\meanUU=U_0\cos k_1x\cos k_1z,
\end{equation}
is driven in the meridional plane
$-\pi/2\leq k_1x\leq0$ and $0\leq k_1z\leq\pi/2$.
In the following we adopt units where $k_1=1$.
The equator is assumed to be at $z=0$ and the outer surface at $x=0$.
The bottom of the convection zone is at $x=-\pi/2$ and
$z=\pi/2$ corresponds to the latitude of around $30^\circ$, i.e.\ where
the surface angular velocity equals the value in the radiative interior.
In the plots below we always display nondimensional combinations by
scaling length with $k_1$ and time with $u_{\rm rms}k_1$.

In this model there is radial shear near the ``bottom'' of what represents
the convection zone and latitudinal shear in the upper parts.
At the level of simplification necessary to isolate fundamentally new
effects, such as the shear--current effect of
Rogachevskii \& Kleeorin (2003, 2004) we have refrained from modeling
the near-surface shear layer.
Furthermore, curvature effects are ignored and no Coriolis force is
included, so the $\OO\times\meanJJ$ effect of R\"adler (1969) is absent.
With nonhelical forcing, the $\meanWW\times\meanJJ$ (where
$\meanWW=\nab\times\meanUU$ is the mean vorticity) is however a possible
effect driving large scale dynamo action.
Nevertheless, we focus on the morphology of the field in the case where the
helicity of the forcing is finite and negative -- consistent with the
conditions in the northern hemisphere of the sun.

We assume an isothermal equation of state with sound speed $c_{\rm s}=\const$,
and solve the continuity, momentum, and induction equations in the form
\begin{equation}
{\DD\ln\rho\over\DD t}=-\nab\cdot\UU,
\end{equation}
\begin{equation}
{\DD\UU\over\DD t}=-c_{\rm s}^2\nab\ln\rho+{\JJ\times\BB\over\rho}
+\ff+\FF_{\rm visc},
\end{equation}
\begin{equation}
{\partial\AAA\over\partial t}=\UU\times\BB-\eta\mu_0\JJ,
\end{equation}
where $\rho$ is density, $\UU$ is velocity, $\JJ=\nab\times\BB/\mu_0$ is
current density, $\BB=\nab\times\AAA$ is magnetic field expressed in terms
of the magnetic vector potential, $\eta$ is the magnetic diffusivity, and
\begin{equation}
\FF_{\rm visc}=\nu\left(\nabla^2\UU+\onethird\nab\nab\cdot\UU
+2\SSSS\cdot\nab\ln\rho\right)
\end{equation}
is the viscous force where $\nu=\mbox{const}$ is the kinematic viscosity, and
${\sf S}_{ij}=\frac{1}{2}(U_{i,j}+U_{j,i})-\frac{1}{3}\delta_{ij}U_{k,k}$
the traceless rate of strain tensor.

We step the equations forward in time by using the {\sc Pencil Code}\footnote{
\url{http://www.nordita.dk/software/pencil-code}}, which is a high order
finite difference code (sixth order in space and third order in time)
for solving the compressible hydromagnetic equations on distributed memory
machines using the Message Passing Interface libraries.
The numerical resolution is $128\times512\times128$ meshpoints.
The boundary conditions are periodic in the $y$ direction and stress-free
in the $x$ and $z$ directions.
For the magnetic field we assume perfect conductor boundary conditions
at what corresponds to the base of the convection zone ($k_1x=-\pi/2$)
and at mid-latitudes ($k_1z=\pi/2$).
At the equator ($z=0$) and at the outer surface ($x=0$) we assume the
magnetic field to be normal to the boundaries, i.e.\ $\BB\times\nnn=0$.
We refer to these boundaries as ``open'', because they permit magnetic
and current helicity fluxes, as opposed to ``closed'' or perfectly conducting
boundaries where $\BB\cdot\nnn=0$ with no helicity fluxes.

The strength of the forcing is chosen such that the typical rms velocity
of the turbulence (without the systematic shear flow) is subsonic with
$u_{\rm rms}/c_{\rm s}=0.1$.
Viscosity and magnetic diffusivity are chosen such that $\nu/\eta=1$ and
$R_{\rm m}=u_{\rm rms}/(\eta k_{\rm f})\approx80$.
In the simulations with negative helical forcing, the turbulent part
of the velocity field is nearly fully helical, i.e.\
$\bra{\oo\cdot\uu}/(k_{\rm f}u_{\rm rms}^2)\approx-1$.
The mean flow is about 5 times stronger than the turbulent flow, i.e.\
$U_0/u_{\rm rms}\approx5$.

\subsection{Growth of the large scale field}

A series of experiments has been performed: with helical forcing of
negative helicity (representative of the northern hemisphere; denoted
by ``$\alpha>0$'' since the resulting electromotive force
would lead to a positive $\alpha$ effect), positive
helicity (mainly for comparison, but it could be representative of the
bottom of the convection zone; denoted by ``$\alpha<0$''),
and without helicity (denoted by ``$\alpha=0$'').
In all cases the kinematic growth rate is about the same
($\dd\ln B_{\rm rms}/\dd t=0.02\,u_{\rm rms}k_{\rm f}$).


\begin{figure}[t!]\begin{center}
\plotone{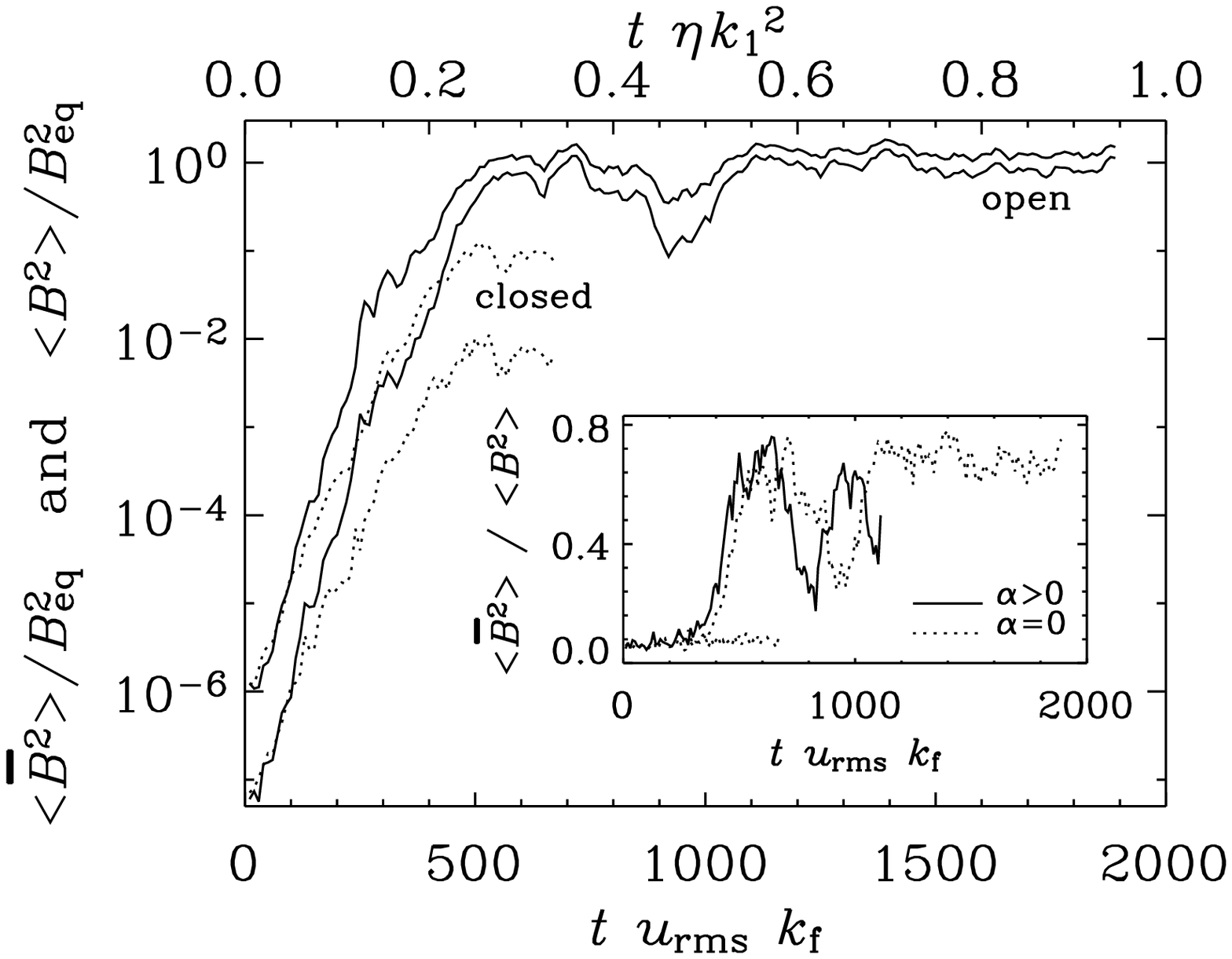}
\end{center}\caption[]{
Evolution of the energies of the total field $\bra{\BB^2}$ and of
the mean field $\bra{\meanBB^2}$, in units of $B_{\rm eq}^2$,
for runs with non-helical forcing
and open or closed boundaries; see the solid and dotted lines, respectively.
The inset shows a comparison of the ratio $\bra{\meanBB^2}/\bra{\BB^2}$
for nonhelical ($\alpha=0$) and helical ($\alpha>0$) runs.
For the nonhelical case the run with closed boundaries is also
shown (dotted line near $\bra{\meanBB^2}/\bra{\BB^2}\approx0.07$).
Note that saturation of the large scale field occurs on a
dynamical time scale; the resistive time scale is given on the
upper abscissa.
}\label{pmean_comp}\end{figure}


In Brandenburg \& Sandin (2004) the effect of boundaries was already
found to be important: when a perfect conductor condition was used at
the equator and at the outer surface the resulting $\alpha$ effect was
found to be suppressed by a factor of $\sim30$.
In the present case we find that near saturation the large scale
field remains well below equipartition (see the dotted line in
\Fig{pmean_comp}).
With open boundary conditions,
near-equipartition field strengths can be achieved
($\meanBB^2/B_{\rm eq}^2\approx0.8$).
Here we define $\meanBB$ as an average over the $y$ direction
(toroidal average).
Volume averages are denoted by angular brackets.

In the presence of finite helicity the result is not greatly affected;
see the inset of \Fig{pmean_comp} where we show that the ratio
$\bra{\meanBB^2}/\bra{\BB^2}$ either varies around 0.5 (for $\alpha>0$
or $\alpha<0$), or that it stays around 0.7 (for $\alpha=0$).
The case $\alpha<0$ is not shown here, but we refer to
Brandenburg et al.\ (2005) for a description of those results.

So far we have not seen reversals of the field.
We note, however, that in principle cycles are possible in this type of
geometry and have indeed be found both in the corresponding mean field
model (Brandenburg \& Sandin 2004).
In the present case the lack of cycles could be connected with the
resulting mean flow that was neglected in the mean field calculations.
[On the average, however, the mean poloidal flow (poleward at the
surface) is less than 2\% of
the total mean flow. By comparison, the mean poloidal field is about
25\% of the total mean field; see \Sec{BipolarRegions}.]

Our main conclusion from these simulations is that large scale dynamo
action can produce equipartition field strengths on a dynamical time
scale, provided the boundaries are open.
The significance of open boundaries is that magnetic and current helicity
can leave the domain, thus preventing the excessive build-up of
small scale magnetic helicity before the large scale field
has saturated.


\begin{figure}[t!]\begin{center}
\plotone{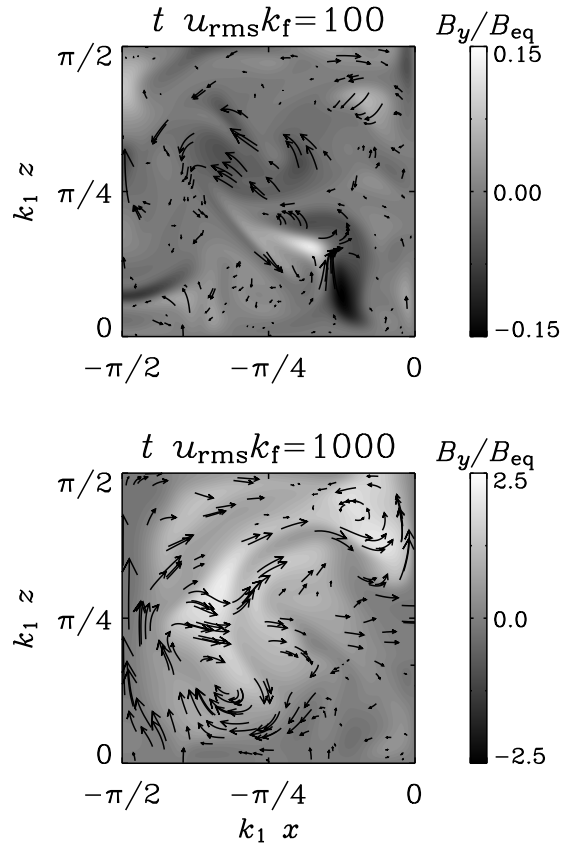}
\end{center}\caption[]{
Snapshots of the magnetic field $\BB$ in the meridional plane
during the kinematic stage ($t=100$ turnover times) and the
saturated stage ($t=1000$ turnover times).
Vectors in the meridional plane are superimposed on a gray scale
representation of the azimuthal field.
The gray scale is symmetric about mid-gray shades, so the absence of
dark shades (e.g.\ in the lower panel) indicates the absence of negative
values.
Note the development of larger scale structures during the saturated
stage with basically unidirectional toroidal field.
}\label{pslice_128b3}\end{figure}


\subsection{How fibril is the field?}

Virtually all dynamo simulations (both small scale and large scale) show
that once the dynamo saturates, the typical length scale of the field
increases, as measured for example by the magnetic Taylor microscale
$\lambda_{\rm M}$, where $\lambda_{\rm M}^2=5\mu_0^2\bra{\JJ^2}/\bra{\BB^2}$;
see Schekochihin et al.\ (2004b) for the case of a forced small scale
dynamo and Brandenburg et al.\ (1996) for the case of a small scale dynamo
in convective turbulence.
In practice this means that the typical scale of the flux structures
increases during the saturation.
However, the orientation of the field is otherwise still random.
The presence of shear together with turbulent diffusion has a strong
tendency to order the field such that it points everywhere in the
same direction.
This tendency has been studied earlier in connection with a completely
random (incoherent) $\alpha$ effect (Vishniac \& Brandenburg 1997).
This tendency is seen in the present simulations as well; see
\Fig{pslice_128b3}, where we show meridional cross-sections of the field during kinematic and saturated phases.
 From the simulations presented here we cannot support the assumption
that in the solar convection zone the field will be highly fibril.

Compared to the sun, there is of course the difference that the
turbulence is not driven by a body force, but by convection.
However, it is not clear that this will make an important difference.
It is also possible that at larger magnetic Reynolds numbers there
will be a stronger tendency to produce intense fibrils.
However, for the present simulations with $128\times512\times128$
meshpoints the viscosity is already as small as possible.
In fact, the magnetic Reynolds number based on the mesh spacing,
$\delta x$, of $R_{\rm m}k_{\rm f}\delta x\approx5$, which is a typical
value that should not be exceeded in these type of simulations.

Once we abandon the rising flux tube picture, we have to think
of other ways to produce bipolar regions with the right tilt angle.
This will be discussed in the next section.


\begin{figure}[t!]\begin{center}
\plotone{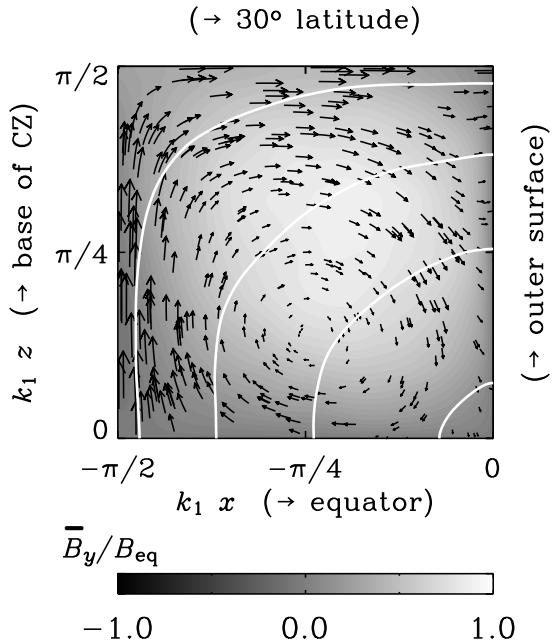}
\end{center}\caption[]{
Gray scale representation of the azimuthally and time {\it averaged} mean
azimuthal field $\meanBB(x,z)$ together with vectors in the meridional plane.
The mean toroidal velocity is shown as white contours.
The projected positions on the sun are labeled on the corresponding axes.
Note the equatorward orientation of the poloidal field near the outer
surface ($x=0$).
As in \Fig{pslice_128b3}, the gray scale is symmetric about mid-gray shades,
so the absence of dark shades indicates that $\meanB_y>0$.
In this run the kinetic helicity is negative, corresponding to $\alpha>0$.
The root mean square of the mean poloidal field is $\sim25\%$ of the
total mean field.
}\label{pbm_128b3}\end{figure}


\begin{figure}[t!]\begin{center}
\plotone{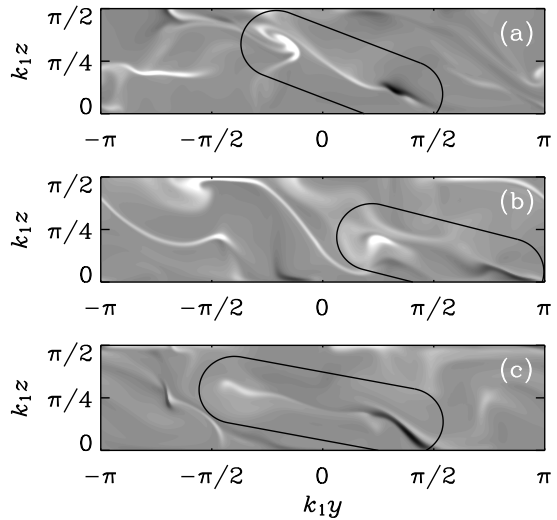}
\end{center}\caption[]{
Magnetograms of the radial field $B_x(y,z,t)$ at the outer surface ($x=0$)
on the northern hemisphere at times $t/\tau=480$, $750$, and $990$,
where $\tau=(u_{\rm rms}k_{\rm f})^{-1}$ is the turnover time,
and $k_{\rm f}$ is the wavenumber corresponding to the
energy-carrying scale of the turbulence.
Light shades correspond to field vectors pointing out of the domain,
and dark shades correspond to vectors pointing into the domain.
The elongated rings highlight the positions of bipolar regions.
Note the clockwise tilt relative to the $y$ (or toroidal) direction,
and the systematic sequence of polarities (white left and dark right)
corresponding to $\meanB_y>0$, which is consistent with \Fig{pbm_128b3}.
}\label{pmagnetogram}\end{figure}


\subsection{Formation of bipolar regions}
\label{BipolarRegions}

In the present simulations, because of shear, most of the field is
in the streamwise direction.
On the open boundaries, on the other hand, the field can only be
normal to the boundary and hence $B_y=0$.
However, because elsewhere in the interior, the field is mostly
toroidal, places with significant normal field ($\BB\cdot\nnn\neq0$)
will be rare.
This is also what is seen in the simulations; see \Fig{pmagnetogram}, were
we show ``magnetograms'' of the normal field on the outer surface, $x=0$.
A meridional cross-section of the azimuthally and temporally averaged field,
$\meanBB$, is shown in \Fig{pbm_128b3}.
Here the identification with a sector in a meridional plane of the sun
is annotated on the axes.
Except near the open boundaries, where $\meanB_y=0$, the mean field is
mostly into the plane ($\meanB_y>0$) and is accompanied by a right-handed
swirl so that $\meanB_z<0$ on the outer surface, $x=0$.

What the magnetograms in \Fig{pmagnetogram} show is basically a gray
background (corresponding to zero field) with only a few patches,
some of which come in pairs.
Often the pairs are connected by a faint ``bridge''.
In all cases the bipolar regions as well as the bridges are inclined
relative to the toroidal direction by an inclination angle that
is primarily determined by the latitudinal shear.
Any cross-stream (i.e.\ latitudinally oriented) field becomes sheared
out and intensifies until the structure disappears.
The polarity depends on the sign of the latitudinal ($z$) component
of the field beneath the surface.
The phases of maximum intensity correspond to times when structures are
most prominent; see \Fig{pmagnetogram}.
Control simulations with opposite sign of helicity confirm that the
inclination is qualitatively unchanged and that the poloidal field
determines the orientation of the polarities, not the toroidal field.
If the tilt was entirely determined by the negative $\alpha$, it would
have produced tilt in the opposite sense.
In the present case of negative kinetic helicity in the northern hemisphere
($\alpha>0$), as in the upper parts of the solar convection zone,
the $\alpha$ effect would produce tilt of the same sign as the shear.
However, as we have seen above, the effect of $\alpha$ on the tilt is
subdominant in the present simulations, where the effect of shear is strong.

The typical separation of the different polarities corresponds to about
1--2 eddy scales ($=2\pi/k_{\rm f}$).
Comparing to the sun, the pressure scale height at $r=0.95R$ is $12\Mm$,
so the mixing length and hence the eddy scale is $\sim20\Mm$.
Thus, the typical size of the active region in the model is $\sim30\Mm$.

We conclude from this section that the tilt of bipolar regions depends
mainly on the latitudinal differential rotation, and that the orientation
of the polarities depends on the orientation of the latitudinal component
of the field rather than its azimuthal component.
The sense of the tilt is thus independent of the sign of $\alpha$.
The simulations suggest that more or less isolated
bipolar regions can emerge in a way that is at least as plausible
as the picture of strong tilted flux tube poking through the surface
from deep underneath.

\section{Conclusions}

It has long been known that sunspots and other magnetic tracers
rotate faster than the photospheric plasma.
Originally, this was taken as evidence that the sun must rotate
faster in the interior (Golub et al.\ 1981).
In fact, it was believed that the angular velocity of sunspots
agrees with the local angular velocity at the depth where the
sunspots are anchored.
Since the mid-eighties this idea became largely discarded on the
grounds that helioseismology began to show angular velocity contours
that are nearly spoke-like and that the only location of radial shear
was the bottom of the convection zone.
 From a dynamo theorist's point of view this result, together
with the already popular idea that the solar dynamo should operate at or
below the bottom of the convection zone (\Tab{ArgumentsSummary}), meant
that at least the issue of the location of the dynamo had been settled.
This picture was in principle quite appealing and it became particularly
attractive in combination with the subsequent finding that the right tilt
angles can be obtained when the fields at the bottom of the convection
zone are of the order of $10^5\G$ (D'Silva \& Choudhuri 1993).

Two important results have emerged since then.
First, downward pumping tends to be a strong effect that can overcome
magnetic buoyancy up to fairly large field strengths.
Second, helioseismology has now revealed the presence of a
near-surface shear layer that is stronger and more prominent
than indicated by the early results of helioseismology.
In the tachocline, by comparison, the radial shear layer is rather
weak at $30^\circ$ latitude (which is where sunspots emerge in the
beginning of the cycle) and extremely strong at the poles (where the
magnetic activity is weak).
It appears that the impact of these findings on the solar dynamo paradigm
ought to be reconsidered.
The purpose of the present paper was therefore to present the arguments
for and against tachocline dynamos versus distributed dynamos that are
possibly strongly affected or shaped by the near-surface shear layer.
The reason we use word ``shaped'' is to indicate that the dynamo is
likely to operate in the entire convection zone, and not only in the
near-surface shear layer.

The idea that sunspots might be anchored at a depth of 0.95 solar radii
has however a problem.
The helioseismologically determined angular velocity at that depth is
still $\sim4\nHz$ slower than that of the very youngest sunspots.
This corresponds to a velocity difference of $\sim20\mpers$.
In principle, if the profile of such enhanced angular velocity is
sufficiently localized, one might argue that the spatial resolution
of helioseismology was still insufficient.
Alternatively, and perhaps more likely, there may still be some not yet
understood mechanism causing newly emerging flux to rotate slightly faster.

An important part of the flux tube paradigm is the idea that strong flux
tubes emerge from deep underneath and form bipolar regions and sunspot
pairs as they reach the surface.
This picture relies entirely on the thin flux tube approximation, which
may have its own difficulties (Dorch \& Nordlund 1998, Wissink et al.\ 2000).
However, as demonstrated in \Sec{Simulations} and through \Fig{pmagnetogram},
a distributed dynamo is quite able to produce bipolar regions with
plausible tilt angles.
As discussed at the end of \Sec{BipolarRegions}, the orientation of
the tilt is controlled by the latitudinal shear.
The orientation of the polarities is determined by the direction of
the poloidal field, and not the azimuthal field.

Since this model lacks convection and stratification, both turbulence
and shear have to be produced by body forces.
Nevertheless, the model is fully self-consistent and not subject to
approximations, such as the thin flux tube approximation.
Obviously, an important next step should be to include convection and
stratification.
Equally important is the implementation of a more realistic outer boundary
condition, possibly allowing for the development of coronal mass ejections
that might be necessary for carrying small scale magnetic and current
helicities away from the dynamo.
Proper modeling of coronal mass ejections might require the use of
spherical geometry.
A lack of magnetic and current helicity fluxes out of the domain would
prevent the dynamo from operating on a dynamical time scale
(Blackman \& Field 2000).
On the other hand, some degree of throttling of the helicity flux might
actually occur in the sun.
This might explain why the solar cycle period tends to be about 10 times
longer than what is suggested by standard mean field
models (K\"ohler 1973) using canonical estimates for
the turbulent diffusivity (Krivodubskii 1984).
Obviously, a reasonably accurate theory of helicity fluxes is required
before this question can be addressed in mean field calculations.

\acknowledgements
I thank the organizers of the program ``Magnetohydrodynamics of Stellar
Interiors'' at the Isaac Newton Institute in Cambridge (UK) for creating
a stimulating environment that led to the present work.
I also thank Antonio Ferriz-Mas, Daniel Gomez, David Moss, Sasha Kosovichev,
Mike Thompson, Jack Thomas, and Nigel Weiss for references, comments,
and inspiring discussions.
The Danish Center for Scientific Computing is acknowledged for granting
time on the Horseshoe cluster.



\vfill\bigskip\noindent\tiny\begin{verbatim}
$Header: /home/brandenb/CVS/tex/mhd/surfdiffrot/ms.tex,v 1.95 2005/02/15 20:30:21 brandenb Exp $
\end{verbatim}

\end{document}